# TRANSPORT ANOMALIES OF THE STRANGE METAL: RESOLUTION BY HIDDEN FERMI LIQUID THEORY

Philip W Anderson, Philip A Casey[*], Physics Dept, Princeton University

ABSTRACT
**The strange metal phase of optimally- and over-doped cuprates exhibits a number of anomalous transport properties—unsaturating linear-T resistivity, distinct relaxation times for Hall angle and resistivity, temperature-dependent anisotropic relaxation times, and a characteristic crossover from supposed Fermi Liquid to linear-T behavior. All receive natural explanations and quantitative fits in terms of the Hidden Fermi Liquid theory.**

## INTRODUCTION

From the very first observations of the properties of the cuprate "high-Tc" superconductors it was seen that the properties of the "normal" metal above Tc were unusual. There are actually two unusual regimes: at lower doping, there develops a "pseudogap" regime which is most plausibly described[i,ii] as a state with BCS pairing but without superconducting order. (At still lower dopings various complex phases with inhomogeneities and/or alternative orderings show up also, but we will consider only homogeneous, translationally symmetric phases.) Near optimal and above (and also for T above the pseudogap regime) there is no evidence of pairing in the normal state but instead a characteristic "strange" metallic behavior extending to very high temperatures and energies. Most obvious is the notorious "linear T" resistivity, sometimes extrapolating at T=0 to 0 or less and persisting in its linearity often to well above the Mott limit. A cleaner characterization, if available, is the "Drude-like" tail of the mid-infrared conductivity. This falls off as a non-integer power of frequency considerably less than the $\omega^{-2}$ of the Drude theory.[iii] Very early a heuristic for the "strange" behavior was developed as the "marginal Fermi Liquid" theory[iv] and that is often used as a descriptive term, but this heuristic does not describe the infrared result correctly, nor any of the further regularities.

[*] This author acknowledges support from an NSERC PGS-D Fellowship.

Often a third regime is postulated, that the state returns to the simple Fermi liquid when overdoped beyond the superconducting dome, and for lower T, a crossover line being drawn up and to the right, starting at the edge of the dome. We will see that this is merely a crossover in the transport properties, and that fundamental properties like the one-particle Green's function remain anomalous according to our theory. No transition to a true Fermi liquid has been observed, in our opinion.

A striking anomaly of the strange phase is the T-dependent Hall effect. It is best described as there existing a relaxation rate for the Larmor precession $1/\tau_H$ distinct from that for the resistivity and more resembling that for a Fermi liquid.[v]

The purpose of the rest of this paper is to show how all of these anomalies follow from the theory of the simplest possible model, the Hubbard model with a strong interaction U and *nothing else*.

THE HIDDEN FERMI LIQUID
The hidden Fermi liquid (HFL, hereafter) theory[vi] depends on the assumption that the Hubbard on-site interaction U is sufficiently strong that it must be renormalized to infinity by the Gros-Rice canonical transformation, leaving behind a superexchange interaction and the kinetic energy projected on the lower Hubbard band. That is, the effective Hamiltonian is

$$H = \sum_{i,j} J_{ij} S_i \cdot S_j + P[\sum_{i,j,\sigma} t_{ij} c^*_{i,\sigma} c_{j,\sigma}] P$$
$$P = \prod_i (1 - n_{i\uparrow} n_{i\downarrow})$$  [1].

This "t-J Hamiltonian" is not simply a convenient alternative to the Hubbard model, it reflects the physical fact that the low-energy states live within a subspace which is overcompletely described by a single full band of electron states, because anti-bound states (doublons) have been ejected out of the top of the band. No convergent perturbative route exists to connect the low states to the original band of the Hubbard model, since they exist within Hilbert spaces of different dimensionality.

It is assumed that in the strange metal region J is too weak, because of competition with kinetic energy[vii] or thermal fluctuations, to cause pair condensation and an anomalous self-energy, and therefore its major effect can be lumped in with that of phonons as a renormalization of the kinetic energy. It will also contribute electron-electron scattering but we do not expect it to be as large as that due to the projection. Therefore the problem reduces to the effect of Gutzwiller projection on the renormalized kinetic energy, represented by a simple Fermi gas, that is, to the second term in H, so we consider the Hamiltonian

$$H_P = P \sum_{i,j,\sigma} t_{ij} c_{i\sigma}^* c_{j\sigma} P = \sum_{i,j,\sigma} t_{ij} \hat{c}_{i\sigma}^* \hat{c}_{j\sigma}$$

$$\hat{c}_{i\sigma} = c_{i\sigma}(1 - n_{i,-\sigma})$$

[2]

where we introduce the projective quasiparticle operators c-hat and c*-hat, which automatically enforce the projection.

The HFL Ansatz is that the projected Hamiltonian [2] operating in the *unprojected* Hilbert space of many-electron wave functions gives one the low-energy spectrum of a Fermi liquid—essentially, that it has a sharp Fermi surface with the usual analyticity properties of the self-energies of the quasiparticles c and c*. The Ansatz can be thought of as the result of a Shankar-style[viii] renormalization but can really be justified only by demonstrating its self-consistency, and by testing to what extent it agrees with experiment; in both respects it seems so far to have passed muster. But the quasiparticles in this Hilbert space are not the true quasiparticles of the physical system: these are the projected quasiparticles which we designate with "hats". We shall hereafter invent the name "pseudoparticle " to describe the objects c and c* which obey Fermi liquid rules because they operate in the full Hilbert space.

The pseudoparticles have renormalized Fermi velocities which can be estimated with the Gutzwiller approximation

$$v_{F,ren} = v_{F,0} g_t \quad with \quad g_t = 2x/(1+x), \quad x \quad being\ the$$
doping percentage.

[3]

They can be expected to have rather large electron-electron scattering proportional to $(k-k_F)^2$. In the one case in which we have accurate information, optimally-doped BISSCO[ix], the coefficient is

$$\Gamma_{ee} = Cv_F^2(k-k_F)^2, C = 3.6 \times 10^{-3} (mev)^{-1} \quad [4]$$

Straightforward phase space considerations would, as observed by Drew,[x] suggest that the coefficient should be of the order 1/W, W being the bandwidth, but in the Hall effect case of interest to him he observed that it was considerably larger, and we also find this: W is of order a few hundred mev rather than a few thousand. A little thought persuades us that this should be the case. The Gutzwiller projection slows the coherent Fermi velocity for an electron with spin near the Fermi surface, but it does not much affect the incoherent motions of bare holes, which are just as rapid as in the unprojected state—for instance, the second moment of the overall spectrum is unaffected.[xi] The quasiparticles are broadened by these incoherent motions proportionately to this second moment, roughly, so one might expect that the broadening would be proportional to $g^{-2}$ or about an order of magnitude larger than the naïve estimate. Thus our hidden Fermi liquid will tend not to be a very good one, in the sense that the coherence of its pseudoparticles lasts only out to 200-300 mev from the Fermi surface. We should also note that there is no reason to expect this scattering mechanism to be anisotropic.

Let us now consider the transport properties of such a system: first the resistivity. As PWA discussed in my book and in related papers[xii], this is complicated by being a two-step process. The momentum is delivered to the system via accelerating the true quasiparticles, ie by displacing their Fermi surface. But the scattering which transfers momentum to the lattice is the $T^2$ umklapp scattering of the pseudoparticles which we have just been discussing. Gutzwiller projection is perfectly translation-invariant, so that the process of decay of true quasiparticles into pseudoparticles is momentum-conserving and cannot lead to resisivity by itself. It acts, instead, as a *bottleneck*, a necessary step which must take place before the true scattering events can operate. (As PWA noted in ref 7, this is actually the same physics which is involved in phonon drag, but I think the "bottleneck" description is clearer.) It is the *slower* of the two processes which will control the rate: they <u>do not add according to Matthiessen's rule but according to its inverse.</u>

In previous work (ref 7) PWA calculated the dissipation due to the quasiparticle decay process by approximating the two-particle Green's function which appears in the response function by the simple product of two one-particle functions, since it should be a good approximation for the quasiparticles to decay independently. In ref 4 and related papers[xiii] we have shown that the form of the single-particle Green's function at absolute zero is the simple expression

$$G(r,t) = G_0(r,t) G^*(t) \begin{Bmatrix} 1 \\ g \end{Bmatrix}$$

$$G^*(t) = t^{-p}; \quad p = (1-x)^2/4 \quad [5]$$

Here the 1 applies on the hole side, the g on the electron. (For finite T, presumably, the jump singularity of the coefficient becomes a Fermi function.) $G_0$ is the pseudoparticle Green's function. In reference [4] we showed how to generalize [5] to finite temperature. $G_0$ follows the conventional rules, while as we pointed out there, the power law in G* was shown by Yuval[xiv] to follow the general rule of being antiperiodic in imaginary time by becoming

$$G^*(t,T) = \left(\frac{\pi T}{\sinh \pi T t}\right)^p \cong e^{-\pi p T t} \, for \, Tt \gg 1 \quad [6]$$

This is the source of the ubiquitous "linear-T" decay. Note that the relaxation rate is isotropic, but the mean free path and therefore the conductivity will have the anisotropy of the Fermi velocity, since the Fermi *momentum* is fairly isotropic.

At high frequencies and high temperatures the $T^2$, $\omega^2$ decay implied by [4] may be assumed to be more rapid than [6] and dissipative processes will be dominated by the power law decay of quasiparticles into pseudoparticles. The most straightforward situation is the infrared conductivity which has long been known to obey a frequency power law [xv],

$$\sigma_{ir}(\omega) \propto (i\omega)^{-1+2p} \quad, \quad [7]$$

which can easily be derived from [5].

Timusk[xvi] has experimentally estimated the dependence of the power 2p on doping, which we show in figure 1; the agreement as to magnitude is good, the dependence on doping a bit slow. But our prediction is within the scatter of the data.

As far as D C resistivity is concerned, [6] accounts for the observed linear dependence on T near optimal doping. The trend with doping is in agreement with the expected $(1-x)^2$ dependence of p, though in order to be quantitative one would need an estimate of the carrier density which is hard to come by.

In the same regime we see the striking phenomenon first observed by Ong[xvii] of a qualitative difference between the relaxation time $\tau$ as estimated from the DC conductivity using $\sigma=ne^2\tau/m$, as opposed to using the Hall angle formula $\Theta_H=\omega_c\tau_H$. The latter shows a conventional Fermi liquid temperature dependence $\propto T^2$, while the resistivity is linear in T as we have just been describing. In the HFL theory this difference is very natural: the Hall angle observed is that of the underlying pseudoparticles of the HFL. The Larmor precession which is caused by the magnetic field does not change relative occupancies and therefore does not disturb the equilibrium between quasiparticles and pseudoparticles: effectively, the magnetic field commutes with Gutzwiller projection. Thus the Hall effect and other magnetic responses—such as the de Haas-van Alfven effect—will be identically those of the HFL, with no bottleneck caused by the decay of the quasiparticles. We have estimated the magnitude of the Hall angle and found that it is reasonably accounted for by our estimates of Drew's W.

The only effect of the strong interaction will be quantitative. As I remarked above, the $T^2$ relaxation rate will be unexpectedly large. We have as yet been unable to get a direct comparison between the relaxation rates as measured from ARPES data and those measured via the Hall effect, because the samples are not comparable; but the general observation of Drew, that the $T^2$ rates are high, seems to be borne out. A more accurate numerical fit would involve a very complete study of the Fermi surface curvature and the anisotropy of the Fermi velocity.

The final topic to take up is the resistivity in the region completely beyond the "dome" which is normally designated as "the Fermi Liquid".[xviii] Indeed, the resistivity at low temperatures seems to obey the $T^2$ law; but we see no reason to suppose that the effects of the strong interaction die out so suddenly. Actually, the resistivity in this region seems to be nicely explained in terms of the bottleneck effect, along with the anisotropy of the HFL conductivity due to the anisotropy of $v_F$.

The temperature dependence of the resistivity, then, is obtained by combining the two conductivities.

$$\sigma_{HFL} = ne^2\tau/m = \frac{e^2}{\hbar}\frac{\hbar^2 k_F^2}{m}\frac{\tau}{\hbar} = \frac{e^2}{\hbar}\frac{E_F W}{T^2} \quad [8]$$

Here we have ignored numerical factors of order 1, realizing that they may be subsumed in the parameter W, the effective bandwidth discussed under equation [4]. Conductivities are 2-dimensional, per single plane, and T is in energy units. The effective conductivity corresponding to the decay process [6] is

$$\sigma_{decay} = \frac{ne^2 v_F \tau}{m v_F} = \frac{e^2 (\hbar k_F v_F)}{\hbar T} = \frac{e^2}{\hbar}\frac{E_F}{T}(v_F/v_{F0}) \quad [9]$$

Here $v_{F0}$ is the maximum Fermi velocity, which gives us an estimate of the overall bandwidth $E_F$; then we make explicit the dependence on Fermi velocity which will indeed vary quite strongly from the diagonal direction to the zone corners (and in the right direction to account for the anisotropy observed by Hussey[xix]).

First we would like to compare the general temperature dependence of the resistivity implied by [8] and [9] with relatively early measurements on overdoped cuprates, where there was no attempt to disentangle the anisotropy (Refs. 18, [xx]). In this case, leaving out the anisotropic Fermi velocity, the resistivity is the universal expression

$$\rho = \frac{\hbar}{e^2 E_F}\frac{T^2}{T+W}(+\rho_{res})$$

$$d(\ln(\rho - \rho_{res}))/d\ln T = 1 + W/(T+W) \quad [10]$$

(some samples show a small residual resistance which we would expect to be simply additive a la Matthiessen's rule, playing no role in the bottleneck.) The fit of the form (10) to the data is quite satisfactory. For instance, in ref 20 (the '92 version) there is a plot of the effective exponent vs T, which for low T, where the data is most accurate, follows the second equation of (10) accurately. Reference 18 fits the data over the entire range with a $T^{3/2}$ power law, which according to [10] should only be approximate; indeed, we get as accurate a fit, except at high T, where the measurement is questionable because of thermal expansion.

Fig 2 shows our fit to the data of ref 18, and Fig 3 the values of the parameters in [8] and [9] obtained from the fit, as a function of doping.

The x-dependence of the parameter W is experimentally even stronger than $x^2$. One aspect which we have not taken into account is that [8] is not the conventional conductivity of the HFL as would appear if the E-field acted directly on it; the three pseudoparticles must recohere into a quasiparticle to interact with the field. Surely this effect works in the right direction.

Reference [19] provides an even more explicit confirmation of our theory. Hussey's equation [3] shows that he is empirically driven to the necessity of adding conductivities [8] and [9], rather than resistivities, but unfortunately not in quite the correct form [10]. His work using angle-dependent magnetoresistance measurements[xxi] has shown experimentally that in the optimal- to overdoped regime, there are two scattering mechanisms for every momentum on the Fermi surface (not "hot" and "cold" spots) with distinct temperature and angle dependences, and as I pointed out above the theory provides precisely those temperature dependences and the correct sign and magnitude for the anisotropy of the linear T term.

In a very recent paper, [xxii] the same group have revisited the doping range of reference 18, but over a very restricted temperature range. They have used a large magnetic field to destroy superconductivity when present so have a lower minimum temperature. Their fitting function is purely empirical and has more parameters to adjust than [10], and in fact we can achieve an equal level of agreement over their limited temperature range (see Fig 4 and parameters in Fig 3).

CONCLUSION
The Hidden Fermi liquid method seems well on the way to providing a complete resolution of the anomalous properties of the "strange metal" phase of the cuprate superconductors. Complex-seeming as they are, these seem to follow from the slightest possible generalization of the conventional Fermi liquid theory of metals, namely the inclusion of the projective constraint made necessary by the existence of strong on-site electron-electron interactions. This simple case, far from being an impenetrable mystery as it is so often pictured to be, should provide the canonical model for more complex examples of strongly interacting electronic systems.

We should acknowledge extensive discussion of the experimental data with N P Ong.

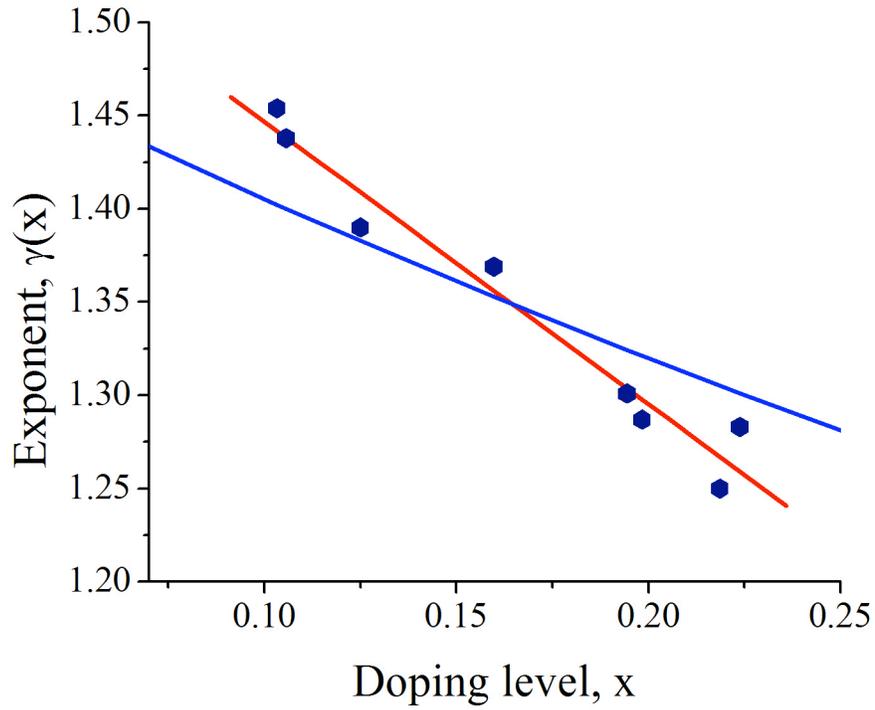

Figure 1. Infrared spectrum exponents for $Bi_2Sr_2CaCu_2O_{8+\delta}$. Data points from ref. 16 with linear best fit of ref. 16 (red line) and predicted value from ref. 23 (blue line). The predicted exponent stems from $\sigma(\omega) = (i\omega)^{-2+\gamma}$ with $\gamma = 1 + 2p$, and p is given in Eq. [5].

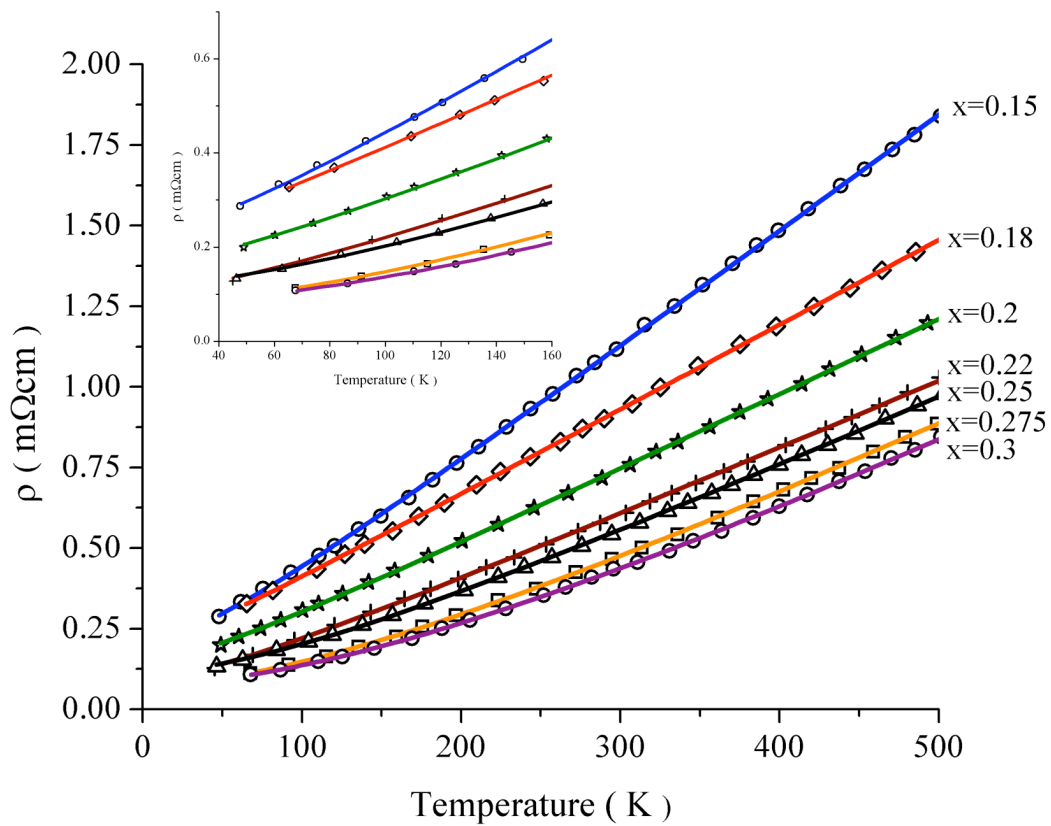

Figure 2. Comparison of the polycrystalline La$_{2-x}$Sr$_x$CuO$_4$ resistivity (data points) extracted from Ref. 18 with the "bottleneck" resistivity form of Eq. [10]. Inset slows the low temperature region in detail.

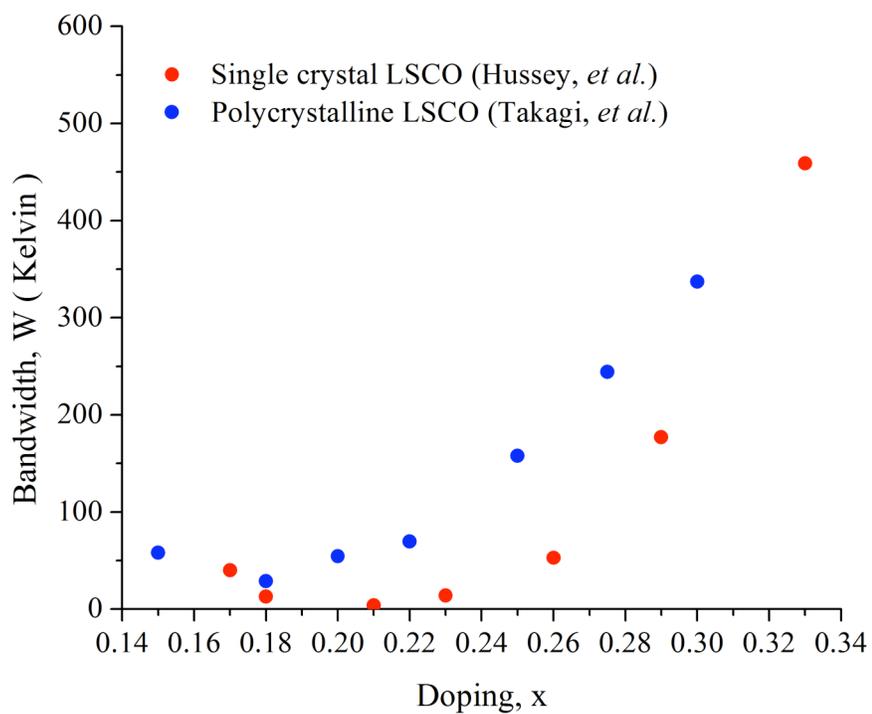

Figure 3a.

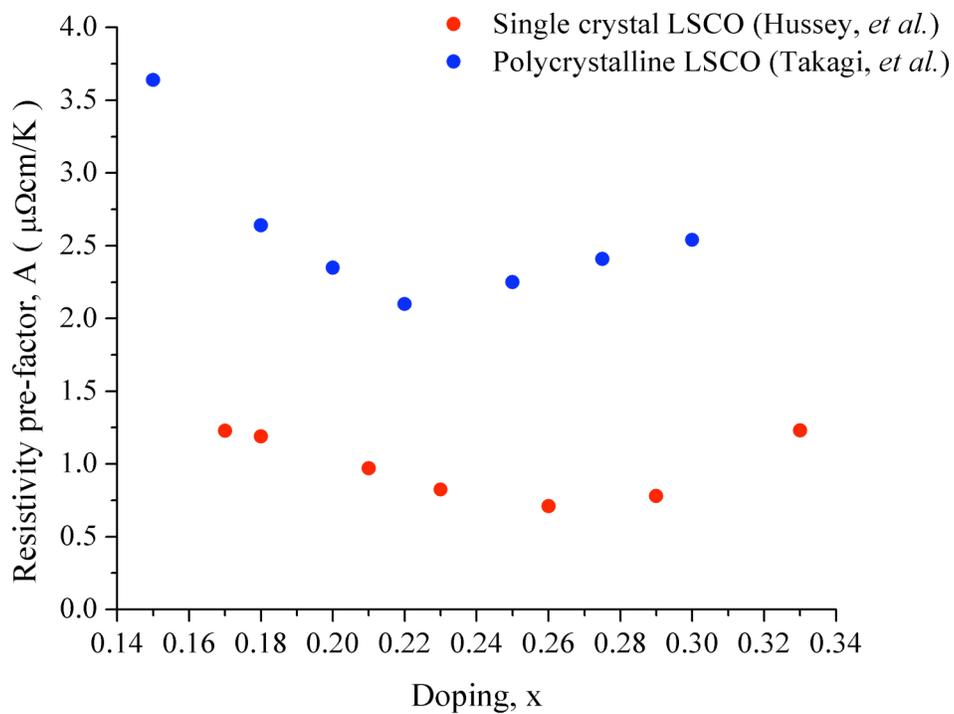

Figure 3b.

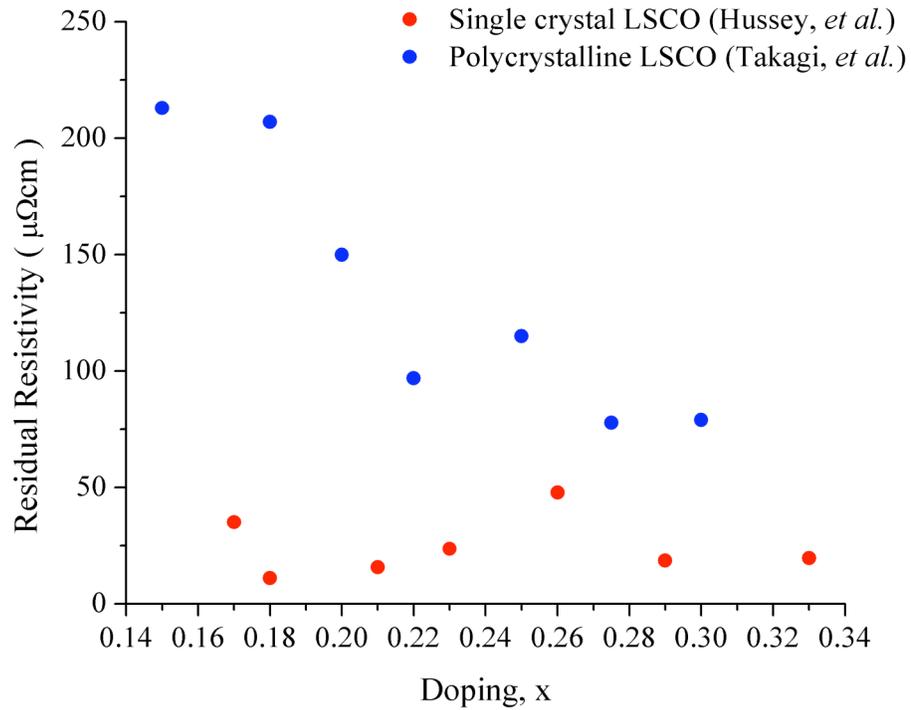

Figure 3c.

Figure 3. Parameters of the "bottleneck" resistivity form of Eq. [10] for comparisons in Fig. 2 and Fig. 4. The three parameters are (a) the bandwidth, (b) a pre-factor for the first term in [10], and (c) the residual resistivity.

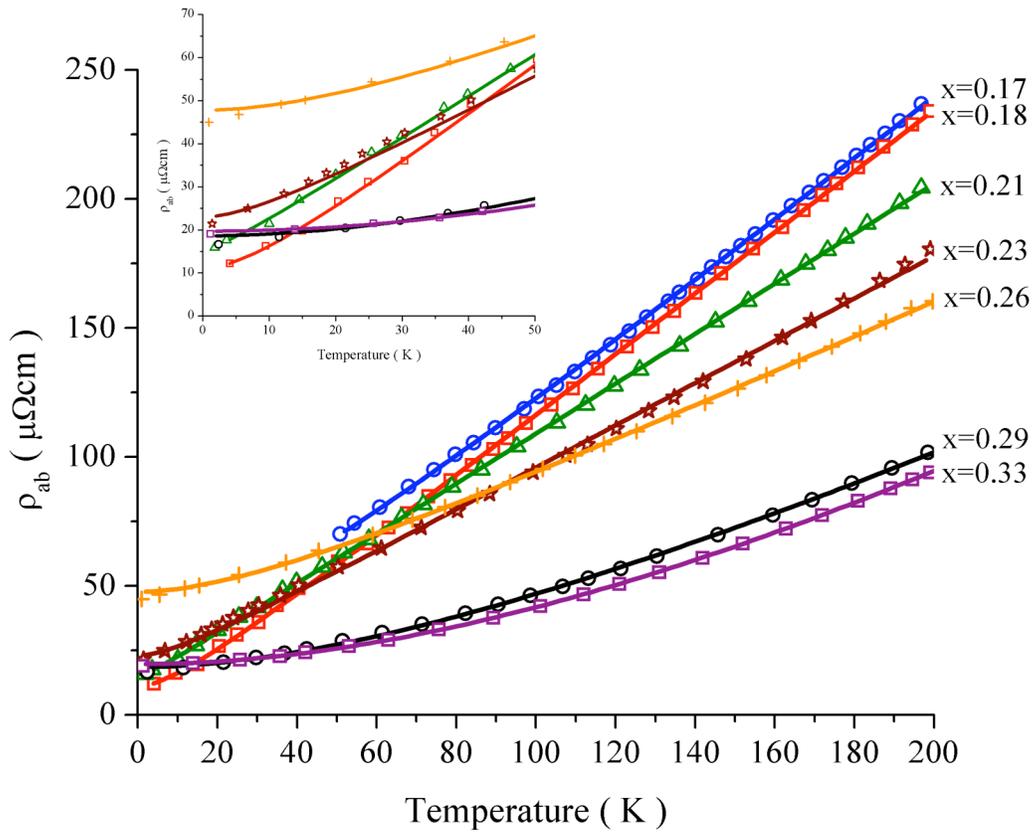

Figure 4. Comparison of the single crystal $La_{2-x}Sr_xCuO_4$ resistivity (data points) extracted from Ref. 22 with the "bottleneck" resistivity form of Eq. [10]. Functional parameters can be found in Fig. 3. Inset slows the low temperature region in detail. The low temperature resistivity data was determined by Hussey, *et al.* by suppressing superconductivity with a large magnetic field and then extrapolating the high field resistivity data to zero field.